\documentclass[prd,onecolumn,showpacs,nofootinbib,aps,prd,longbibliography]{revtex4-1}
\usepackage{amsmath}
\usepackage{tensor}
\usepackage{latexsym}
\usepackage{amsfonts}
\usepackage{graphicx}
\usepackage{mathrsfs}
\usepackage{hyperref}
\usepackage{mathrsfs}
\usepackage{accents}
\usepackage{color}

\makeatletter
\def\@bibdataout@aps{%
\immediate\write\@bibdataout{%
@CONTROL{%
apsrev41Control%
\longbibliography@sw{%
    ,author="08",editor="1",pages="1",title="0",year="1"%
    }{%
    ,author="08",editor="1",pages="1",title="",year="1"%
    }%
  }%
}%
\if@filesw \immediate \write \@auxout {\string \citation {apsrev41Control}}\fi 
}
\makeatother

\begin{document}

\title{Gravitational waveforms from the  inspiral of compact binaries in the Brans-Dicke theory in
an expanding Universe}

\author{Tan Liu$^1$}\email{lewton@mail.ustc.edu.cn}
\author{Yan Wang$^1$}\email{ywang12@hust.edu.cn}
\author{Wen Zhao$^{2,3}$}\email{wzhao7@ustc.edu.cn}

\affiliation{$^1$MOE Key Laboratory of Fundamental Physical Quantities Measurements, Hubei Key
Laboratory of Gravitation and Quantum Physics, PGMF, Department of Astronomy, and School of Physics, Huazhong
University of Science and Technology, Wuhan 430074, China}
\affiliation{$^2$CAS Key Laboratory for Researches in Galaxies and Cosmology, Department of Astronomy, \\
University of Science and Technology of China, Chinese Academy of Sciences, Hefei, Anhui 230026, China}
\affiliation{$^3$School of Astronomy and Space Science, University of Science and Technology of China, Hefei 230026, China}

\begin{abstract}
In modified gravity theories, such as the Brans-Dicke theory, the background evolution of the Universe and the perturbation around it are different from that in general relativity. Therefore, the gravitational waveforms used to study standard sirens in these theories should be modified. The modifications of the waveforms can be classified into two categories: wave generation effects and wave propagation effects. Hitherto, the waveforms used to study standard sirens in the modified gravity theories  incorporate only the wave propagation effects and ignore the wave generation effects; while the waveforms focusing on the wave generation effects, such as the post-Newtonian waveforms, do not incorporate the wave propagation effects and cannot be directly applied to the sources with non-negligible redshifts in the study of standard sirens. In this work, we construct the consistent waveforms for standard sirens in the Brans-Dicke theory. The wave generation effects include the emission of the scalar breathing polarization $h_b$ and the corrections to the tensor polarizations $h_+$ and $h_\times$; the wave propagation effect is the modification of the luminosity distance for the gravitational waveforms. Using the consistent waveforms, we analyze the parameter estimation biases due to the ignorance of the wave generation effects. Considering the observations by the Einstein Telescope, we find that the ratio of the theoretical bias to the statistical error of the redshifted chirp mass is two orders of magnitude larger than that of the source distance. For black hole-neutron star binary systems like GW191219, the theoretical bias of the redshifted chirp mass can be several times larger than the statistical error.
\end{abstract}

\maketitle

\section{Introduction}
The Hubble constant describes the present expansion rate of the Universe \cite{2005pfc..book.M}. The recent local measurement of the Hubble constant from the Ia supernovae explosions is a factor of 9\%  larger than the large scale measurement from the cosmic microwave background data. This discrepancy is  $5\sigma$ and is called the Hubble tension  \cite{2021arXiv211204510R}. In addition to these methods, gravitational waves (GWs) can be used to determine the Hubble constant. Since the GWs emitted by the compact binaries are well modeled and the measured waveforms can be used to determine their cosmological distances. Such binaries are called standard sirens \cite{Schutz1986,Holz2005} which are the GW analog with electromagnetic standard candles. Combined with the redshift of the GW sources, the Hubble constant can be measured by the observation of the standard sirens. The detection of the binary neutron star merger event GW170817 in both GWs and electromagnetic waves is the first realization of the measurement of the Hubble constant by the standard siren \cite{GWHubble}. Standard sirens can provide new insight into the solution of the Hubble tension \cite{PhysRevLett.125.201301}. For a review of the standard sirens, please see \cite{ZhaoSiren}.

% Gravitational waves (GWs) can be used to study the expanding history of the Universe. Since the GWs emitted by compact binaries are well modeled and the measured waveforms can be used to determine their cosmological distances. Such binaries are called standard sirens \cite{Schutz1986,Holz2005} which are the GW analog with electromagnetic standard candles. Combined with the redshift of the GW sources, the Hubble constant can be measured by the observation of standard sirens. The detection of the binary neutron star merger GW170817 in both GWs and electromagnetic waves is the first realization of the measurement of the Hubble constant by standard siren \cite{GWHubble}.
% In addition to this method, there are others methods to determine the Hubble constant. The recent local measurement from supernovae is inconsistent  with  the large scale measurement from the cosmic microwave background data  at $5\sigma$ level \cite{2021arXiv211204510R}. Because of this discrepancy, the Hubble crisis arises. Standard sirens can put new perspectives on trying to solve this crisis \cite{PhysRevLett.125.201301}. For a review of standard sirens, please see \cite{ZhaoSiren}.

The Hubble tension can also be solved by the modified gravity theories \cite{Valentino2021a}. It has been shown that  the tension between the local and large scale measurement of the Hubble constant can be smoothed out if the observational data are interpreted in the Brans-Dicke theory rather than in general relativity (GR) \cite{2019ApJ...886L...6S}. The Brans-Dicke theory is a scalar-tensor theory which is an extension of GR with a massless scalar field \cite{PhysRev.124.925,*Dicke1959,*Dicke1962}.  A large and growing body of literature has investigated  GWs in scalar-tensor theories \cite{PhysRevD.104.124017,Hou2020,PhysRevD.104.104010,PhysRevD.87.104029,Maselli2022,PhysRevD.102.124035,2022arXiv220913749H}. The gravitational waveforms emitted by inspiralling compact  binaries  have been calculated systematically using the post-Newtonian (PN) method \cite{Oliynyk2009,Poisson2014} when the gravitational fields are weak and the orbital velocities are small. The leading order Newtonian quadrupole waveforms have been obtained in \cite{PhysRevD.50.6058}. Then, the higher PN order contributions to the waveforms were calculated in different approaches \cite{Will-BDn,PhysRevD.89.084014,Lang-BD,PhysRevD.94.084003,Bernard2022}. In these PN waveforms, the binary system is modeled as two point particles moving along  circular orbits. In recent years, the eccentricity of the orbit \cite{PhysRevD.100.124032}, the spins of the components \cite{PhysRevD.102.024053,Brax2021}, and the tidal interaction between the components \cite{Bernard2020} have been incorporated  into the waveforms. However,  these waveforms and other waveforms focusing on wave generation effects \cite{Maselli2022,PhysRevD.87.081506,PhysRevD.85.102003} cannot be used to study standard sirens in cosmological distances  with non-negligible redshifts because they are calculated on the Minkowski background.

Focusing on the Brans-Dicke theory as an example, the present work serves as a first step to coherently construct gravitational waveforms  from compact binaries in the expanding Universe in the modified gravity theories.
Compared with GR, the modified gravity theories can predict different expansion history of the background Friedmann-Lema\^{i}tre-Robertson-Walker (FLRW) universe \cite{CLIFTON20121}  and  change the gravitational waveforms which are  perturbations around the background \cite{Barack_2019}. The modifications of the waveforms can be classified into two categories: wave generation effects \cite{PhysRevD.98.084042,*PhysRevD.101.109902} and wave propagation effects \cite{PhysRevD.97.104037,*PhysRevD.97.104038,*PhysRevD.99.104038}. It is well known that GWs have two tensor polarizations ($h_+$ and $h_\times$) in GR \cite{misner1973gravitation}. During wave generation, the modified gravity can induce extra polarization(s) as well as amplitude  and phase corrections to the tensor polarizations \cite{PhysRevD.98.084042,*PhysRevD.101.109902}. When the waves propagate through the Universe, the modified gravity can also produce amplitude and phase correction to the tensor polarizations \cite{PhysRevD.97.104037,*PhysRevD.97.104038,*PhysRevD.99.104038,PhysRevD.101.024002}. However, all previous studies on the standard sirens in the modified gravity have ignored the wave generation effects and used inconsistent gravitational waveforms \cite{PhysRevD.99.083504,PhysRevD.102.044036,PhysRevD.103.064075,Baker_2021,Belgacem_2019,Dalang_2019,PhysRevD.98.023510,PhysRevD.100.044041,PhysRevLett.124.061101,Belgacem2019,PhysRevD.103.104059,Amendola2018}. For instance, see Eq. (14) in \cite{PhysRevD.99.083504} and Eq. (12) in \cite{PhysRevD.103.104059}.

To construct consistent waveforms which consider both the wave generation and wave propagation effects, we follow Thorne's suggestion \cite{Thorne1987,RevModPhys.52.299} that divides the wave zone where the waves propagate into local wave zone and distant wave zone. The local wave zone is near the GW source and the effect of the background curvature of the universe is negligible. Thus, the background spacetime can be seen as flat in the local wave zone and the PN method can be used to solve the field equations to obtain the GW waveforms. Then, using the geometric optics approximation \cite{PhysRev.166.1263,PhysRev.166.1272,PhysRevD.102.044036,PhysRevD.103.064075,2022arXiv220900795K,Garoffolo_2020}, the waveforms in the local wave zone propagate into the distant wave zone in which the curvature of the universe is important.  In this way, the waveforms in the distant wave zone contain both wave generation and wave propagation effects.

In this paper, we construct the consistent waveforms on the background of an  expanding Universe in the Brans-Dicke theory based on the Newtonian quadrupole waveforms \cite{PhysRevD.50.6058,PhysRevD.102.124035}. In particular, we obtain the time domain waveforms of two tensor polarizations  and one scalar polarization (breathing polarization $h_b$) in the distant wave zone (cf. Eqs. \eqref{otz}-\eqref{hb2z}). We incorporate the modifications that originate  from both wave generation and propagation. These waveforms are new results. As mentioned above, the wave generation effects include amplitude correction, phase correction, and additional polarization to the ones in GR; but in the Brans-Dicke theory, during wave propagation, there is only amplitude correction, and this amplitude correction appears as a modification of the electromagnetic luminosity distance. The modification of the luminosity distance in the tensor polarizations is consistent with previous studies \cite{PhysRevD.99.083504,PhysRevD.102.044036,PhysRevD.103.064075,Baker_2021,Belgacem_2019,Dalang_2019,PhysRevD.98.023510,PhysRevD.100.044041,PhysRevLett.124.061101,Belgacem2019,PhysRevD.103.104059,Amendola2018,PhysRevD.97.104066}, while the modification of the luminosity distance in the breathing polarization is a new result. 
Furthermore, we analyze the parameter estimation biases due to the inconsistent waveforms which ignore the wave generation effects. Considering the observation of the GWs from the  black hole-neutron star (BH-NS) binaries by Einstein Telescope \cite{Punturo2010} and setting the source parameters to be that of the BH-NS candidates detected by LIGO/Virgo \cite{2021arXiv211103606T}, we find that the inconsistent waveforms can bias the redshifted chirp mass significantly while have negligible influence on the measurement of the distance when the Brans-Dicke coupling constant saturates the bound imposed by the Cassini spacecraft \cite{PhysRevD.85.064041}.  For BH-NS systems like GW191219, the theoretical bias of the redshifted chirp mass can be several times larger than the statistical error.

The paper is organized as follows. In section \ref{wavelocal}, we review the PN waveforms in the local wave zone in the Brans-Dicke theory. In section \ref{geo}, we construct the geometric optics equations for wave propagation in curved background spacetime in the Brans-Dicke theory. In section \ref{proFLRW}, we apply the geometric  optics approximation to propagate the waves through the FLRW universe and obtain the waveforms in the distant wave zone. In section \ref{estbias}, we estimate the theoretical bias due to the inconsistent waveforms. In section \ref{condis}, we summarize  the results and discuss the prospects  for future research.  

For the metric, Riemann, and Ricci tensors, we follow the conventions of Misner, Thorne, and Wheeler \cite{misner1973gravitation}. We adopt the unit $c=1$, with $c$ being the speed of light. We do not set the gravitational constant $G$ equal to 1, since the effective gravitational constant  will vary over the history of the universe in this theory.

\section{Gravitational waveforms in the local wave zone }\label{wavelocal}
The space around the  GW source can be divided into three regions by  two length scales $r_\text{I}$ and $r_\text{O}$:
a near zone $(r\lesssim r_\text{I})$, a local wave zone $(r_\text{I}\lesssim r\lesssim r_\text{O})$, and a distant wave zone $(r\gtrsim r_\text{O})$.  
The inner boundary of the local wave zone $r_{\rm I}$ is much larger than the GW wavelength so that the source's gravity is weak in this region. The outer boundary of the local wave zone $r_{\rm O}$ is large enough that this region contains many wavelengths; but not so large that the background curvature of the universe influences the propagation of the GWs. 
Therefore, the GWs can be regarded as propagating in flat spacetime in the local wave zone. This zone is the matching region for the problem of wave generation and wave propagation \footnote{For GWs in the millihertz to kilohertz band, $r_{\rm I}$ and $r_{\rm O}$ can be chosen to be 100 wavelengths and 1000 wavelengths, respectively. Under this choice, the local wave zone is only a subset of the local wave zone proposed by Thorne, but is sufficient for the application in this paper. For more information about the local wave zone, please see section 9.3.1 in \cite{Thorne1987} and section III in \cite{RevModPhys.52.299}. }. 

In this section, we review the wave generation problem and present the tensor and scalar waveforms in the local wave zone. Detailed derivation can be found in  \cite{PhysRevD.50.6058,PhysRevD.102.124035}. The action of the Brans-Dicke theory is \cite{PhysRevD.85.064041}
\begin{equation}\label{action}
S=\frac{1}{16\pi}\int d^4 x\sqrt{-g}\left[\phi R-\frac{\omega(\phi)}{\phi} \partial_\mu\phi\partial^\mu\phi+M\right]+S_m\left[g_{\mu\nu},\Psi_m\right],
\end{equation}
where $g\equiv \det g_{\mu\nu}$ and   $\omega(\phi)$ is the coupling function  \cite{PhysRevLett.70.2220}. The  constant  $M$ is associated with the effective cosmological constant and $\Psi_m$ denotes the matter fields collectively. 
The field equations are \cite{PhysRevD.85.064041}

\begin{equation}\label{teq}
R_{\mu\nu}-\frac12 g_{\mu\nu}R-\frac12 \frac{M}{\phi}g_{\mu\nu}=\frac{8\pi}{\phi}T_{\mu\nu}+\frac{\omega(\phi)}{\phi^2}(\phi_{,\mu}\phi_{,\nu}-\frac12 g_{\mu\nu}\phi_{,\alpha}\phi^{,\alpha})+\frac{1}{\phi}(\nabla_\mu\nabla_\nu-g_{\mu\nu} \square)\phi
\end{equation}

\begin{equation}\label{seq}
\square\phi-\frac{2M}{2\omega(\phi)+3}=\frac{8\pi}{2\omega(\phi)+3}(T-2\phi~ T')-\frac{\omega'}{2\omega(\phi)+3}\phi_{,\alpha}\phi^{,\alpha}
\end{equation}
where $'\equiv\frac{d}{d\phi}$, the commas denote partial derivative,  and $\square\equiv\nabla_\nu\nabla^\nu$. The covariant derivative $\nabla_\nu$ satisfies $\nabla_\nu g_{\alpha\beta}=0$.
$T_{\mu\nu}$ is the stress-energy tensor of matter and $T=g^{\mu\nu}T_{\mu\nu}$.

In order to solve the wave generation problem inside the outer radius  $(r\lesssim r_\text{O})$, the field equations should be expanded around the Minkowski background  $\eta_{\mu\nu} dx^\mu dx^\nu=-dt^2+dr^2+r^2(d\iota^2+\sin^2\iota~ d\sigma^2)$ and the scalar background $\bar{\phi}$,
\begin{equation}
\phi= \bar{\phi}+\varphi, \qquad g_{\mu\nu}=\eta_{\mu\nu}+h_{\mu\nu},
\end{equation}
\begin{equation}\label{thetamunu}
\theta_{\mu\nu}\equiv h_{\mu\nu}-\frac12 h \eta_{\mu\nu}-\frac{\varphi}{\bar{\phi}}\eta_{\mu\nu},
\end{equation}
where $|h_{\mu\nu}|\ll 1$ and $|\varphi/\bar{\phi}|\ll 1$. When we deal with the wave generation problem, the scalar background $\bar{\phi}$ can be viewed as a constant, although it will evolve with the expansion of the Universe. Imposing the harmonic gauge
\begin{equation}
\theta^{\mu\nu}_{\phantom{12},\mu}=0,
\end{equation} 
the field equations become
\begin{equation}\label{weak-t}
\square_\eta \theta_{\mu\nu}=-16\pi G \tau_{\mu\nu},
\end{equation}
\begin{equation}\label{weakscalar}
\square_\eta \varphi=-16\pi \mathbb{S},
\end{equation}
where $\square_\eta=\eta^{\mu\nu}\partial_\mu\partial_\nu$ and  $\tau_{\mu\nu}=T_{\mu\nu}+t_{\mu\nu}$. The source term $t_{\mu\nu}$ denotes the nonlinear perturbations collectively. The effective gravitational constant is
\begin{equation}
G \equiv \frac{1}{\bar{\phi}}.
\end{equation}
The  cosmological constant is discarded when dealing with the wave generation. For the explicit expression of the source $\mathbb{S}$ of the scalar wave equation, see Eq. (18) in \cite{PhysRevD.102.124035}.

Consider the waves emitted by a binary system on a quasicircular orbit with orbital frequency $\omega_e$\footnote{The subscript stands for `emission'. The observed orbital frequency will be redshifted.}. When the velocity of the binary system is slow and the gravitational field is weak, the above wave equations can be solved by the PN method. The waveforms emitted by the binary system will depend on the sensitivities of the binary bodies, defined by
\begin{equation}
s_A\equiv \frac{d \ln m_A(\phi)}{d \ln \phi}\Big|_{\bar{\phi}}.
\end{equation}
Since the scalar field can influence the gravitational binding energy of the compact object, the inertial mass $m_A(\phi)$ of object $A$ is a function of the scalar field \cite{1975ApJ...196L..59E}. The sensitivity of a black hole is $\frac12$  \cite{1975ApJ...196L..59E} and the typical value of the sensitivity of a neutron star is about $0.2\sim 0.3$ \cite{1992ApJ...393..685Z,PhysRevD.85.064041}.

The tensor waves emitted by this binary system are given by \cite{PhysRevD.102.124035}
\begin{equation}
\theta^{\rm TT}_{\mu\nu}=h_+ e^+_{\mu\nu}+h_\times e^\times_{\mu\nu},
\end{equation}
where `TT' denotes the transverse and traceless part. In the local wave zone, the waveforms of the two tensor polarizations at the Newtonian quadrupole order are 
\begin{equation}\label{hphc}
h_+ = -4 \frac{(GM_c)^{5/3}}{r}(\tilde{g}\omega_e)^{2/3}\frac{1+\cos^2\iota}{2}\cos\psi\Big|_{t_r},
\end{equation}
\begin{equation}
h_\times = -4 \frac{(GM_c)^{5/3}}{r}(\tilde{g}\omega_e)^{2/3}\cos\iota ~\sin\psi\Big|_{t_r},
\end{equation}
where the chirp mass $M_c$ is given by
\begin{equation}
M_c \equiv \mu^{3/5} m^{2/5}
\end{equation}
with $m=m_1+m_2$, $\mu=m_1 m_2/m$, and $m_A = m_A(\bar{\phi})$. 
\begin{equation}\label{gtilde}
\tilde{g}\equiv 1+\alpha(1-2s_1)(1-2s_2),\qquad \alpha\equiv \frac{1}{2\omega(\bar{\phi})+3}.
\end{equation}
The factor $\tilde{g}$ originates from the modification of  Kepler's law \cite{PhysRevD.102.124035}.

The coordinates are adapted so that the binary system is at the origin $r=0$ and $\iota$ is the angle between the line of sight and the angular momentum of the binary system. The phase of the tensor wave is given by
\begin{equation}\label{phase}
\psi(t_r)=2\int^{t_r} \omega_e(t')dt',
\end{equation}
where $\omega_e(t)$ is the  orbital frequency of binary system at time $t$ and the retarded time is
\begin{equation}\label{ret}
t_r = t-r.
\end{equation}
In the local wave zone, the polarization tensors are
\begin{equation}\label{epec}
e^{\mu\nu}_+ = e_{(\iota)}^\mu e_{(\iota)}^\nu-e_{(\sigma)}^\mu e_{(\sigma)}^\nu,
\qquad e^{\mu\nu}_\times = e_{(\iota)}^\mu e_{(\sigma)}^\nu + e_{(\sigma)}^\mu e_{(\iota)}^\nu,
\end{equation}
where
\begin{equation}\label{basis}
e_{(\iota)} = \frac{1}{r}\frac{\partial}{\partial\iota}, \qquad 
e_{(\sigma)} = \frac{1}{r\sin\iota}\frac{\partial}{\partial\sigma}.
\end{equation}

In the local wave zone, the waveforms of the  scalar wave to the Newtonian quadruple order  are 
\begin{equation}\label{swave}
\varphi = \phi_1+\phi_2,
\end{equation}
where
\begin{equation}
\phi_1 = -2 \alpha ~\frac{\mu}{r} ~2 \mathcal{S}(G\tilde{g}m\omega_e)^{1/3}\sin\iota\cos(\frac12\psi)\Big|_{t_r},
\end{equation}
\begin{equation}
\phi_2 = 2 \alpha ~\frac{\mu}{r} ~\Gamma (G\tilde{g}m\omega_e)^{2/3}\sin^2\iota\cos(\psi)\Big|_{t_r},
\end{equation}
with
\begin{align}
\begin{split}
\mathcal{S}&\equiv s_1-s_2,\\
\Gamma &\equiv \frac{(1-2s_1)m_2+(1-2s_2)m_1}{m}.
\end{split}
\end{align}
$\phi_1$ and $\phi_2$ represent the scalar dipole and quadrupole radiation, respectively.

Scalar and tensor waves will carry away the orbital energy of the binary system, and the orbital frequency will increase with time. Using the PN method, the time derivative of the orbital frequency has been obtained \cite{PhysRevD.50.6058,PhysRevD.102.124035}
\begin{equation}\label{orb-fre}
\frac{d \omega_e}{d t_r} = \frac{96}{5}K(GM_c)^\frac53\omega_e^\frac{11}{3}+K_1 G\mu \omega_e^3
\end{equation}
with  
\begin{equation}\label{kk1}
K\equiv \tilde{g}^\frac23(1+\frac16\alpha\Gamma^2),\qquad K_1\equiv 4\alpha\mathcal{S}^2.
\end{equation}
The evolution of the phase $\psi$ can be obtained by using the above equations.

\section{Wave propagation and geometric optics approximation}\label{geo}
The last section has solved the wave generation problem inside the outer radius $(r\lesssim r_{\rm O})$.
This section will establish the equations governing the wave propagation outside the inner radius $(r\gtrsim r_{\rm I})$.   
In this region, the tensor and scalar waves can be seen as small ripples in a  curved, slowly changing background. The frequency of the ripples is high compared to the time scale of background changes, but their amplitudes are small.
Therefore, outside the inner radius, tensor and scalar waves can be viewed as  linear perturbations around the background and the high frequency approximation, i.e., geometric optics approximation, can be used to solve the perturbation equations.
% We picture a gravitational wave as a small ripple in
% the geometry of space-time running through a highly
% curved, slowly changing background. The frequency of
% the ripple is high but its amplitude quite small,
\subsection{Linear perturbations}
Consider the perturbations over an arbitrary background,
\begin{equation}
g_{\mu\nu}=\bar{g}_{\mu\nu}+h_{\mu\nu},\qquad \phi = \bar{\phi}+\varphi.
\end{equation}
The overhead bar denotes the background quantity.
Let us study the perturbations $h_{\mu\nu}$ and $\varphi$ via the field equations \eqref{teq} and \eqref{seq} linearized about the background $\bar{g}_{\mu\nu}$ and $\bar{\phi}$.
Note that the scalar background $\bar{\phi}$ is viewed as a constant when dealing with the wave generation in the previous section. But it can evolve in space and time when we consider the wave propagation problem on much larger scales.

From the field equations \eqref{teq} and \eqref{seq}, the linear perturbation equations are
\begin{align}
\begin{split}\label{tp}
&\bar{\phi}(-\frac12\bar{\square}\hat{h}_{\mu\nu}+\tensor{\hat{h}}{_{\beta(\mu|}^\beta_\nu_)}-\frac12\bar{g}_{\mu\nu}\tensor{\hat{h}}{_{\alpha\beta|}^{\alpha\beta}})-\bar{\nabla}_\mu\bar{\nabla}_\nu\varphi+\bar{g}_{\mu\nu}\bar{\square}\varphi-\bar{F}(2\bar{\phi}_{|(\mu}\varphi_{|\nu)}-\bar{g}_{\mu\nu}\bar{\phi}^{|\alpha}\varphi_{|\alpha})\\
&+\frac12\bar{\phi}^{|\beta}(2\hat{h}_{\beta(\mu|\nu)}-\hat{h}_{\mu\nu|\beta}-2\bar{g}_{\mu\nu}
\tensor{\hat{h}}{_{\beta\rho|}^\rho}-\bar{g}_{\beta(\mu}\hat{h}_{|\nu)}+\frac12\bar{g}_{\mu\nu}\hat{h}_{|\beta})=0,
\end{split}
\end{align}
\begin{equation}\label{sp}
\bar{\square}\varphi-(\frac{1}{\bar{\phi}}-\frac{2\bar{\omega}'}{2\bar{\omega}+3})\bar{\phi}_{|\alpha}\varphi^{|\alpha}=0.
\end{equation}
Here $\bar{\nabla}_\mu$ and the slash $|$ denote the covariant derivative such that $\bar{\nabla}_\mu\bar{g}_{\alpha\beta}=\bar{g}_{\alpha\beta|\mu}=0$ and $\bar{\square}\equiv\bar{\nabla}_\mu\bar{\nabla}^\mu$. The background metric $\bar{g}^{\mu\nu}$ is used to raise the indices. The hat indicates the trace-reversed part of a tensor, e.g., $\hat{h}_{\mu\nu}=h_{\mu\nu}-\frac12\bar{g}^{\mu\nu}\bar{g}^{\alpha\beta}h_{\alpha\beta}$ and $\hat{h}=\bar{g}^{\alpha\beta}\hat{h}_{\alpha\beta}$.
The function $F$ is given by $F\equiv\frac{\omega(\phi)}{\phi}$. Indices placed between round brackets are symmetrized, e.g., 
$\hat{h}_{\beta(\mu|\nu)}=\frac12(\hat{h}_{\beta\mu|\nu}+\hat{h}_{\beta\nu|\mu})$.

Note that terms that do not contain derivatives with respect to the perturbation fields are discarded. Since we focus on the high frequency wave solutions of the perturbations, a term is larger when it contains more derivatives. It will be shown in the following subsection that the above perturbation equations are adequate for the accuracy required in this paper.
Because the interaction between matter and GWs is weak, the perturbation of the stress-energy tensor $\delta T_{\mu\nu}$ induced by GWs is also ignored \cite{1983grr..proc....1T}. The tensor perturbation equation \eqref{tp} is consistent with Eqs. (A9)-(A14) and (A21)-(A27) in \cite{PhysRevD.103.064075}\footnote{There is a typo in Eq. (A28) in \cite{PhysRevD.103.064075}, where 
$-\bar{g}_{\mu\nu}\tensor{\hat{h}}{^{\rho\sigma}_{;\sigma}}$ should be replaced by $-2\bar{g}_{\mu\nu}\tensor{\hat{h}}{^{\rho\sigma}_{;\sigma}}$.}. The scalar perturbation equation \eqref{sp} is consistent with Eqs. (A50)-(A53) and (A57)-(A64) in \cite{PhysRevD.103.064075}.

The second derivatives of $\varphi$ in the tensor perturbation equation \eqref{tp} can be eliminated by introducing the eigentensor perturbation \cite{PhysRevD.103.064075}
\begin{equation}
\theta_{\mu\nu}=\hat{h}_{\mu\nu}-\bar{g}_{\mu\nu}\frac{\varphi}{\bar{\phi}}.
\end{equation} 
Imposing the harmonic gauge 
\begin{equation}\label{harmonic}
\tensor{\theta}{_{\mu\nu|}^\mu}=0,
\end{equation}
the tensor perturbation equation \eqref{tp} is further simplified to 
\begin{align}
\begin{split}\label{tp2}
&-\frac12\bar{\phi}~\bar{\square}\theta_{\mu\nu}+\frac12\bar{\phi}^{|\beta}(\theta_{\beta\mu|\nu}+
\theta_{\beta\nu|\mu}-\theta_{\mu\nu|\beta})-\frac14(\bar{\phi}_{|\mu}\theta_{|\nu}+\bar{\phi}_{|\nu}\theta_{|\mu}-\bar{g}_{\mu\nu}\theta_{|\beta}\bar{\phi}^{|\beta})\\
&-(\frac{3}{2\bar{\phi}}+\bar{F})(\bar{\phi}_{|\mu}\varphi_{|\nu}+\bar{\phi}_{|\nu}\varphi_{|\mu}-\bar{g}_{\mu\nu}\varphi_{|\beta}\bar{\phi}^{|\beta})=0,
\end{split}
\end{align}
where $\theta\equiv\bar{g}^{\mu\nu}\theta_{\mu\nu}$. The above equation is consistent with Eqs. 
(A54)-(A56), (A65)-(A67), and (A73)-(A74) in \cite{PhysRevD.103.064075} \footnote{The right hand side of Eq. (A68) in \cite{PhysRevD.103.064075} misses a term $-2\tensor{K}{_{\mu\nu}^{\rho\sigma(\alpha\beta)}}\hat{C}_{\rho\sigma;\beta}$.}.

\subsection{Geometric optics approximation}
Now we consider the high frequency wave ansatz
\begin{equation}\label{ansatz}
\varphi = \Re[\Phi e^{i v}],\qquad \theta_{\mu\nu}=\Re[\Theta_{\mu\nu} e^{i \psi}]
\end{equation}
of the perturbation equations \eqref{sp} and \eqref{tp2}. Here $\Phi$ and $\Theta_{\mu\nu}$ denote the complex amplitudes of the waves, and $v$ and $\psi$ are their phases. $\Re$ denotes the real part of the argument. Since the waves satisfy the geometric optics approximation outside the inner radius, their phases vary much more rapidly than their amplitudes and the background fields,
\begin{equation}
\partial v,~\partial \psi \gg \partial \ln|\Phi|,~\partial \ln |\Theta_{\mu\nu}|,~\partial \ln|\bar{\phi}|,~\partial \ln|\bar{g}_{\mu\nu}|.
\end{equation}
Therefore, the terms in the perturbation equations can be sorted according to their power of the wave vectors
\begin{equation}
q_\alpha \equiv -\partial_\alpha v,\qquad k_\alpha \equiv -\partial_\alpha \psi.
\end{equation}
Note that, for both the tensor and scalar waves, we  consider only one frequency component in this section, although there are multiple frequency components in realistic situations (cf. Eq. \eqref{swave}).  Because  different frequency components are uncoupled  during propagation, they can be treated separately \cite{PhysRevD.103.064075}. It can also be seen from Eq. \eqref{swave} that the phase $v$ of the scalar wave is proportional to the phase $\psi$ of the tensor wave. Consequently, $q_\alpha$ and $k_\alpha$ are in parallel.

Substituting the wave ansatz \eqref{ansatz} into Eq. \eqref{tp2} yields
\begin{align}
\begin{split}
&-\frac12\bar{\phi}[-k^\alpha k_\alpha \Theta_{\mu\nu}-2i k_\alpha\bar{\nabla}^\alpha\Theta_{\mu\nu}
-i\bar{\nabla}^\alpha k_\alpha \Theta_{\mu\nu}]e^{i \psi}+\frac12\bar{\phi}^{|\beta}[-ik_\nu \Theta_{\beta\mu}-i k_\mu \Theta_{\beta\nu}+i k_\beta \Theta_{\mu\nu}]e^{i\psi}\\
&-\frac14[-i k_\nu\bar{\phi}_{|\mu} -i k_\mu\bar{\phi}_{|\nu} +i\bar{g}_{\mu\nu}k_\beta\bar{\phi}^{|\beta} ]\Theta e^{i\psi}+(\frac{3}{2\bar{\phi}}+\bar{F})[-i\bar{g}_{\mu\nu}q_\alpha\bar{\phi}^{|\alpha} +i q_\mu\bar{\phi}_{|\nu} 
+i q_\nu\bar{\phi}_{|\mu}]\Phi e^{iv}=0,
\end{split}
\end{align}
where $\Theta\equiv \bar{g}^{\mu\nu}\Theta_{\mu\nu}$ and we have discarded terms without $k_\alpha$ or $q_\alpha$.

To the leading order of the above equation, we have 
\begin{equation}\label{tnull}
k^\alpha k_\alpha =0.
\end{equation}
Therefore, 
\begin{equation}\label{tgeo}
k^\mu\bar{\nabla}_\mu k_\nu= -k^\mu \bar{\nabla}_\mu\bar{\nabla}_\nu \psi =  -k^\mu \bar{\nabla}_\nu\bar{\nabla}_\mu \psi
 =\frac12 \bar{\nabla}_\nu(k^\mu k_\mu) =0.
\end{equation}
It suggests  that the tensor waves travel along the null geodesics of the background spacetime.

To the next-to-leading order, we have
\begin{align}
\begin{split}\label{tamp}
&-\frac12\bar{\phi}[2 k_\alpha\bar{\nabla}^\alpha\Theta_{\mu\nu}
+\bar{\nabla}^\alpha k_\alpha \Theta_{\mu\nu}]+\frac12\bar{\phi}^{|\beta}[k_\nu \Theta_{\beta\mu}+ k_\mu \Theta_{\beta\nu}- k_\beta \Theta_{\mu\nu}]\\
&-\frac14[k_\nu\bar{\phi}_{|\mu} + k_\mu\bar{\phi}_{|\nu} -\bar{g}_{\mu\nu}k_\beta\bar{\phi}^{|\beta} ]\Theta +(\frac{3}{2\bar{\phi}}+\bar{F})[\bar{g}_{\mu\nu}q_\alpha\bar{\phi}^{|\alpha} - q_\mu\bar{\phi}_{|\nu} 
- q_\nu\bar{\phi}_{|\mu}]\Phi e^{i(v-\psi)}=0.
\end{split}
\end{align}
This equation governs the evolution of the amplitude of the tensor waves.
It is obvious that the  terms without derivatives with respect to the perturbation fields do not contribute to the geometric optics equations \eqref{tnull}-\eqref{tamp}. Thus, these terms can be safely discarded in the tensor perturbation equation \eqref{tp}. This argument also applies to the scalar perturbation equation \eqref{sp}.

Substituting the wave ansatz \eqref{ansatz} into the scalar perturbation equation \eqref{sp},
the leading order term yields
\begin{equation}\label{qq}
q^\alpha q_\alpha =0,
\end{equation}
and the next-to-leading order terms yields
\begin{equation}\label{qphi}
2 q^\mu\bar{\nabla}_\mu\Phi+ \bar{\nabla}^\mu q_\mu \Phi + (\frac{2\bar{\omega}'}{2\bar{\omega}+3}-\frac{1}{\bar{\phi}})\bar{\phi}^{|\alpha}q_\alpha \Phi=0.
\end{equation}
This is the scalar amplitude evolution equation. The scalar waves also travel along null geodesics.

\section{Propagation through the FLRW universe}\label{proFLRW}
In the previous sections, we have obtained the tensor and scalar waveforms emitted by a binary system in the local wave zone $(r_\text{I}\lesssim r\lesssim r_\text{O})$, and the geometric optics equations for wave propagation from the local wave zone to the distant wave zone $(r\gtrsim r_\text{O})$. In this section, we apply the geometrical optics equations to the situation where the background spacetime is the FLRW universe \footnote{Note that the inhomogeneous distribution of matter can distort the gravitational waveforms \cite{PhysRevD.103.123021,PhysRevD.95.044029}. We restrict to the homogeneous and isotropic FLRW background for simplicity.}. The waveforms in the local wave zone are used as the initial conditions of the geometric optics equations  for propagation through the FLRW universe to obtain the waveforms in the distant wave zone. 

\subsection{Evolution of the phases and amplitudes}
The FLRW metric is 
\begin{equation}\label{flrw}
\bar{g}_{\mu\nu}dx^\mu dx^\nu =a^2(\xi)[-d\xi^2+d\chi^2+\Sigma^2(d\iota^2+\sin^2\iota ~d\sigma^2)]
\end{equation}
where $\Sigma \equiv \chi \text{ for a spatially flat universe}, \sin\chi \text{ for a closed universe}, \sinh\chi \text{ for an open universe}$. We recall that the coordinates are adapted so that the binary system is at the origin $\chi=0$ and $\iota$ is the angle between the line of sight and the angular momentum of the binary system. The relation between the conformal time $\xi$ and the physical time $t$ is given by \cite{2005pfc..book.M}
\begin{equation}\label{pt}
t=\int_{\xi_I}^\xi a(\xi') d \xi',
\end{equation}
where $\xi_I$ corresponds to the beginning of the universe. Inside the outer radius, when we consider the GW generation, the scale factor $a(\xi)$ can be seen as a constant, and the FLRW metric \eqref{flrw} reduces to the Minkowski metric $\eta_{\mu\nu}$. Then
$\{t,r,\iota,\sigma\}$ become the flat, spherical coordinates  where 
\begin{equation}
r= a\chi.
\end{equation}

To have a better understanding of the wave propagation problem, we rewrite the waveform of $h_+$ in the local wave zone as
\begin{equation}
h_+\equiv \Re[A_+e^{i\psi}]
\end{equation}
where 
\begin{equation}\label{Aplus}
A_+ = -4 \frac{(GM_c)^{5/3}}{r}(\tilde{g}\omega_e)^{2/3}\frac{1+\cos^2\iota}{2}
\end{equation}
and 
\begin{equation}\label{ptr}
\psi(t_r)=-2\left[\frac{t_c-t_r}{5GM_c}\right]^{5/8}K^{-3/8}+\frac{5}{56}\left[\frac{t_c-t_r}{5Gm}\right]^{7/8}K_1\eta^{-1/8}
+\psi_c.
\end{equation}
The phase $\psi$ is obtained by integrating Eq. \eqref{orb-fre} twice. $\psi_c\equiv \psi(t_c)$ and $t_c$ is the retarded time such that $\omega_e(t_c)=\infty$. $\eta\equiv \mu/m$ is the symmetric mass ratio. The retarded time $t_r$ is given by Eq. \eqref{ret} and $K$ and $K_1$ are given by Eq. \eqref{kk1}.

First we study the evolution of the phase $\psi$ in the (local and distant) wave zone.
From the geometric optics equation \eqref{tnull}, we have
\begin{equation}
\frac{d\psi}{d\lambda}=0
\end{equation}
where $\lambda$ is the affine parameter of the null rays along which the tensor waves propagate. It shows that the phase $\psi$ is constant along each ray. In the local wave zone, Eq. \eqref{ptr} of the phase satisfies this equation. However, Eq. \eqref{ptr} cannot be directly used in the distant wave zone, since Eq. \eqref{ret} of the retarded time  applies only in the local wave zone. Therefore, we need to extend the null rays to the distant wave zone and obtain the expression of the retarded time in this zone. 

Outside the inner radius, when we consider GW propagation, the rays along which the waves propagate have constant $\iota$ and $\sigma$ due to the spherical symmetry of the FLRW metric. The difference between the conformal time and the radial coordinate, $\xi_e\equiv\xi-\chi$, is also constant along each ray since the rays are null. Here, $\xi_e$ is the conformal time at which the ray is emitted. Therefore, each ray can be characterized by $\iota$, $\sigma$, and $\xi_e$. Since the retarded time on each ray is the physical time of emission, we have\cite{1983grr..proc....1T,Thorne2017a}
\begin{equation}\label{ret2}
t_r = \int_{\xi_I}^{\xi_e} a(\xi') d\xi' .
\end{equation}
In the local wave zone, the retarded time becomes
\begin{equation}
t_r = \int_{\xi_I}^\xi a(\xi') d \xi' - \int_{\xi-\chi}^\xi a(\xi') d \xi' = t-a\chi = t-r
\end{equation}
which is consistent with Eq. \eqref{ret}. Substituting Eq. \eqref{ret2} into \eqref{ptr} yields the phase throughout the local and distant wave zone.

Next, we study the evolution of the amplitude in the (local and distant) wave zone and take $A_+$ as an example.

In the wave zone, in order to propagate the tensor waves, we need the following orthonormal basis vectors 
\begin{equation}
e_{(\iota)}=\frac{1}{a\Sigma}\frac{\partial}{\partial \iota},\qquad e_{(\sigma)} = \frac{1}{a\Sigma\sin\iota}\frac{\partial}{\partial \sigma}
\end{equation}
and the polarization tensors $e^{\mu\nu}_+$ and $e^{\mu\nu}_\times$ which are defined by Eq. \eqref{epec}. 
It can be shown by straightforward calculation  that they are parallel-transported along the rays
\begin{equation}\label{parallel}
k^\mu\bar{\nabla}_\mu e_{(\iota)}^\alpha=k^\mu\bar{\nabla}_\mu e_{(\sigma)}^\alpha=0,
\qquad k^\mu\bar{\nabla}_\mu e^{\mu\nu}_+=k^\mu\bar{\nabla}_\mu e^{\mu\nu}_\times=0.
\end{equation}
The polarization tensors are transverse to the tensor wave vector
\begin{equation}\label{ktrans}
k_\mu e^{\mu\nu}_+=k_\mu e^{\mu\nu}_\times=0.
\end{equation}
This is compatible with the harmonic gauge condition \eqref{harmonic}. Since $q_\mu$ is parallel to $k_\mu$, the polarization tensors are also transverse to the scalar wave vector
\begin{equation}\label{qtrans}
q_\mu e^{\mu\nu}_+=q_\mu e^{\mu\nu}_\times=0.
\end{equation}
In the local wave zone, the basis vectors reduce to Eq. \eqref{basis}. 

Dalang et al. show that the observational effects of the tensor waves are determined by the transverse-traceless part of the tensor waves  \cite{PhysRevD.102.044036,PhysRevD.103.064075}.
Therefore we focus on the evolution of the amplitude of $\theta^{\rm TT}_{\mu\nu}=h_+ e^+_{\mu\nu}+h_\times e^\times_{\mu\nu}$. Contracting $e^+_{\mu\nu}$ with Eq. \eqref{tamp} yields
\begin{equation}\label{aplus}
\bar{\phi}[2k^\alpha \bar{\nabla}_\alpha A_+ + A_+ \bar{\nabla}^\alpha k_\alpha ] + \bar{\phi}^{|\beta} k_\beta A_+=0
\end{equation}
where we have used the parallel-transportation condition \eqref{parallel} and the transverse relations \eqref{ktrans} and \eqref{qtrans}.

To solve the amplitude evolution equation \eqref{aplus}, we shall  dedicate a few lines to the property of the bundle of null geodesics. 
Consider a bundle of null geodesics emerging from the GW source. The cross section of this bundle extends from $\iota_0$ to $\iota_0+\delta\iota$ and from $\sigma_0$ to $\sigma_0+\delta\sigma$ and its area is $a^2\Sigma^2\sin \iota_0 ~\delta \iota ~\delta \sigma$ which satisfies \cite{misner1973gravitation,Poisson2004} 
\begin{equation}
k^\mu\bar{\nabla}_\mu(a^2\Sigma^2\sin \iota_0 ~\delta \iota ~\delta \sigma) = (a^2\Sigma^2\sin \iota_0 ~\delta \iota ~\delta \sigma) ~ \bar{\nabla}_\mu k^\mu.
\end{equation}
Since $\iota$ and $\sigma$ are constants along each geodesic, we have
\begin{equation}\label{nablak}
\bar{\nabla}_\mu k^\mu =k^\mu\bar{\nabla}_\mu \ln(a^2\Sigma^2).
\end{equation}
Substituting this equation into Eq. \eqref{aplus} yields
\begin{equation}\label{aplus1}
k^\mu\bar{\nabla}_\mu \ln(A_+ a\Sigma~\bar{\phi}^\frac12)=\frac{d}{d\lambda}\ln(A_+ a\Sigma~\bar{\phi}^\frac12)=0.
\end{equation}
It shows that the combination of $(A_+ a\Sigma~\bar{\phi}^\frac12)$ is constant along each ray \footnote{Using the amplitude evolution equation, we have $\bar{\nabla}_\mu ( |A_+|^2 \bar{\phi} ~k^\mu)=0$
which can be interpreted as the conservation law of the graviton number, where $ |A_+|^2 \bar{\phi}$ is proportional to the graviton number density \cite{PhysRevD.102.044036,PhysRev.166.1272}.}. This is a first order differential equation. Its solution can be determined by using $A_+$ in the local wave zone, Eq. \eqref{Aplus}, as the initial condition. The evolution of the scale factor $a$ and the scalar background $\bar{\phi}$ is determined by the background evolution equations (see section \ref{be}).
The solution takes the form
\begin{equation}\label{Ap}
A_+=\frac{g_+(t_r,\iota,\sigma)}{a(t)\Sigma\sqrt{\bar{\phi}(t)}}.
\end{equation}
Here $g_+(t_r,\iota,\sigma)$ is constant along each null ray and its value in the distant wave zone can be determined by its value in the local wave zone.  
In the local wave zone, using Eq. \eqref{Aplus}, we have \footnote{Actually, $g_+(t_r,\iota,\sigma)$ has a phase factor $e^{-2i\sigma}$ (cf. Eq. (27.71) in \cite{Thorne2017a}). For a fixed observer, this phase factor can be absorbed by the constant $\psi_c$ in Eq. \eqref{ptr}. Therefore, we ignore this phase factor in this paper. }
\begin{equation}
g_+(t_r,\iota,\sigma)=A_+r \sqrt{\bar{\phi}}= -4\sqrt{\bar{\phi}}\tilde{g}^\frac23(GM_c)^\frac53\omega_e^\frac23\frac{1+\cos^2\iota}{2}.
\end{equation}

Therefore, in the local and distant wave zone, the waveform of $h_+$ polarization is 
\begin{equation}\label{hp}
h_+ =-\frac{4}{a(t)\Sigma\sqrt{\bar{\phi}(t)}}\frac{1+\cos^2\iota}{2}\left\{\sqrt{\bar{\phi}}\tilde{g}^\frac23(GM_c)^\frac53\omega_e^\frac23\cos\psi\right\}_{t_r}.
\end{equation}
Here the subscript $t_r$ denotes that the quantities inside the curly brackets take the value at the retarded time $t_r$.
Similarly, the waveform of $h_\times$ polarization is 
\begin{equation}\label{hc}
h_\times =-\frac{4}{a(t)\Sigma\sqrt{\bar{\phi}(t)}}\cos\iota\left\{\sqrt{\bar{\phi}}\tilde{g}^\frac23(GM_c)^\frac53\omega_e^\frac23\sin\psi\right\}_{t_r},
\end{equation}

Using the scalar wave \eqref{swave} in the local wave zone as the initial condition, the scalar wave geometric optics equations \eqref{qq} and \eqref{qphi} can be solved in the same way.
Since the breathing polarization $h_b$ is proportional to the scalar perturbation \cite{PhysRevD.102.124035}, its waveform is
\begin{equation}\label{hb}
 h_b = -\frac{\varphi}{\bar{\phi}(t)}=h_{b1}+h_{b2}
 \end{equation} 
with
\begin{equation}\label{hb1}
h_{b1}=\frac{\sin \iota}{a(t)\Sigma\sqrt{\bar{\phi}(t)(2\bar{\omega}(t)+3)}}\left\{\frac{2\mu}{\sqrt{\bar{\phi}(2\bar{\omega}+3)}}2\mathcal{S}(G\tilde{g}m\omega_e)^\frac13\cos(\frac12\psi)\right\}_{t_r}
\end{equation}
and 
\begin{equation}\label{hb2}
h_{b2}=-\frac{\sin^2 \iota}{a(t)\Sigma\sqrt{\bar{\phi}(t)(2\bar{\omega}(t)+3)}}\left\{\frac{2\mu}{\sqrt{\bar{\phi}(2\bar{\omega}+3)}}\Gamma(G\tilde{g}m\omega_e)^\frac23\cos(\psi)\right\}_{t_r}.
\end{equation}
We recall that $\bar{\omega}\equiv\omega(\bar{\phi})$.

\subsection{Cosmological redshift effect}
The waveforms in the previous subsection contain some parameters that are not directly observable by GW observation, such as the scale factor $a$, the scalar background $\bar{\phi}$, and the orbital frequency $\omega_e$. We need to rewrite these waveforms in terms of parameters that are directly observable.

The frequency of the tensor waves emitted at the retarded time $t_r$ is $2\omega_e(t_r)$, while the received  frequency will be redshifted by the expanding universe. The cosmologically redshifted wave frequency is
\begin{equation}
2\omega = \frac{\partial \psi}{\partial t} =\frac{\partial \psi}{\partial t_r}\frac{\partial t_r}{\partial \xi}/\frac{\partial t}{\partial \xi} = 2 \omega_e(t_r)\frac{a(t_r)}{a(t)}=\frac{2\omega_e(t_r)}{1+z}
\end{equation}
with $\omega$  the redshifted orbital frequency and $z$  the cosmological redshift of the GW source. We have used Eqs. \eqref{pt} and \eqref{ret2} to obtain the above relation.
The time derivative of the redshifted orbital frequency is 
\begin{equation}
\frac{d \omega}{dt}= \frac{1}{(1+z)^2}\frac{d\omega_e}{d t_r}=\frac{96}{5}K(GM_z)^\frac53\omega^\frac{11}{3}+K_1 G\mu_z \omega^3,
\end{equation}
where $M_z \equiv M_c(1+z)$ and $\mu_z \equiv \mu (1+z)$ are the redshifted chirp mass and redshifted reduced mass, respectively. 
Therefore,
\begin{equation}\label{otz}
\omega(t)=\frac{1}{GM_z}\left[\frac{256}{5}K\frac{t_z-t}{GM_z}\right]^{-3/8}-\frac{K_1}{64Gm_z}\left[\frac{256}{5}\frac{t_z-t}{Gm_z\eta}\right ]^{-1/8},
\end{equation}
where $t_z$ is the physical time such that $\omega(t_z)=\infty$ and $m_z \equiv m(1+z)$.
It can be seen that the evolution of $\omega(t)$ is directly related to the redshifted masses, similar to that in GR. In terms of the redshifted masses, the phase \eqref{ptr} can be rewritten as
\begin{equation}\label{ptz}
\psi(t)=-2\left[\frac{t_z-t}{5GM_z}\right]^{5/8}K^{-3/8}+\frac{5}{56}\left[\frac{t_z-t}{5Gm_z}\right]^{7/8}K_1\eta^{-1/8}
+\psi_c.
\end{equation}

For ease of application, it is necessary to express the waveforms in terms of the quantities which have taken into account the cosmological redshift effect. The waveforms \eqref{hp}-\eqref{hb2} become
\begin{equation}\label{hpz} 
h_+ =-\frac{4}{D_G}\tilde{g}^\frac23(GM_z)^\frac53\omega^\frac23\frac{1+\cos^2\iota}{2}\cos\psi,
\end{equation}
\begin{equation}\label{hcz} 
h_\times =-\frac{4}{D_G}\tilde{g}^\frac23(GM_z)^\frac53\omega^\frac23\cos\iota\sin\psi,
\end{equation}
and
\begin{equation}\label{hbz}
 h_b =h_{b1}+h_{b2}
 \end{equation} 
with
\begin{equation}\label{hb1z}
h_{b1}=2\alpha\frac{G\mu_z}{D_S}2\mathcal{S}(G\tilde{g}m_z\omega)^\frac13\sin \iota\cos(\frac12\psi)
\end{equation}
and 
\begin{equation}\label{hb2z}
h_{b2}=-2\alpha\frac{ G\mu_z}{D_S}\Gamma(G\tilde{g}m_z\omega)^\frac23\sin^2 \iota\cos(\psi).
\end{equation}
Here $G$, $\alpha$, and $\tilde{g}$ take the value at the retarded time $t_r$. We recall that $G \equiv \frac{1}{\bar{\phi}}$, $ \alpha\equiv \frac{1}{2\omega(\bar{\phi})+3}$, and $\tilde{g}\equiv 1+\alpha(1-2s_1)(1-2s_2) $. We have introduced two cosmological distances, the gravitational distance
\begin{equation}\label{dg}
D_G \equiv a(t)\Sigma(1+z)\sqrt{\frac{\bar{\phi}(t)}{\bar{\phi}(t_r)}}
\end{equation}
and the scalar distance
\begin{equation}\label{ds}
D_S \equiv a(t)\Sigma(1+z)\sqrt{\frac{\bar{\phi}(t)\alpha(t_r)}{\bar{\phi}(t_r)\alpha(t)}}.
\end{equation}
These two distances are proportional to the electromagnetic luminosity distance $D_L \equiv a(t)\Sigma(1+z)$. The difference between $D_G$ and $D_L$ originates from the time variation of the effective Planck mass. The gravitational distance \eqref{dg} is consistent with previous studies, e.g., Eq. (97) in \cite{PhysRevD.103.064075} and Eq. (14) in \cite{PhysRevD.99.083504}. However, the scalar distance \eqref{ds} is different from the scalar distance defined by Eq. (103) in Dalang et al. \cite{PhysRevD.103.064075}, since we focus on the breathing polarization $h_b$ which is directly observable but they discuss the scalar perturbation\footnote{In addition, there is a typo in Eq. (103) in \cite{PhysRevD.103.064075},  which should be corrected as $ D_S = \sqrt{\frac{N(\bar{\varphi}_o,\bar{X}_o)}{N(\bar{\varphi}_s,\bar{X}_s)}} D_L \ne D_G$.}. 

The above time domain waveforms \eqref{otz}-\eqref{hb2z} are  new results.
Compared with the previous studies in modified gravity theories, our technical improvement is that we reveal the initial conditions of the geometric optics equations explicitly and use the initial conditions to solve these equations. To  our best knowledge, the previous studies on the waveforms in the local wave zone have never attempted to propagate the waveforms to the distant wave zone; while the previous studies on the wave propagation effects have never attempted to find the initial conditions by solving the wave generation problem.

Compared with the gravitational waveforms in GR (Eqs. (4.191)-(4.195) in \cite{maggiore2008gravitational}), the waveforms \eqref{otz}-\eqref{hb2z} have amplitude correction $\tilde{g}^\frac23$, phase correction due to $K$ and $K_1$, an extra polarization $h_b$, and different cosmological distances $D_G$ and $D_S$. Here $\tilde{g}^\frac23$, $K$ and $K_1$, and $h_b$ represent wave generation effects, while $D_G$ and $D_S$ are wave propagation effects.
Note that the factor $\tilde{g}^\frac23$ is not absorbed into the definition of gravitational distance $D_G$, since $\tilde{g}$ defined by Eq. \eqref{gtilde} depends on the properties of the binary system, besides its location in the universe.
We recall that $\tilde{g}$ corresponds to the modification of Kepler's law \cite{PhysRevD.102.124035}.

\subsection{Background evolution}\label{be}
Substituting the FLRW metric and the scalar background into the field equations \eqref{teq} and \eqref{seq} yields the background evolution equations \cite{2019ApJ...886L...6S}
\begin{equation}
3H^2 +3 H \frac{~\dot{\bar{\phi}}}{\bar{\phi}}-\frac12\omega(\bar{\phi})\left(\frac{~\dot{\bar{\phi}}}{\bar{\phi}}\right)^2+\frac12 \frac{M}{\bar{\phi}}=\frac{8\pi}{\bar{\phi}}\rho_{\rm T},
\end{equation}
\begin{equation}
2\dot{H}+3H^2 +\frac{~\ddot{\bar{\phi}}}{\bar{\phi}} +2 H \frac{~\dot{\bar{\phi}}}{\bar{\phi}}+\frac12\omega(\bar{\phi})\left(\frac{~\dot{\bar{\phi}}}{\bar{\phi}}\right)^2+\frac12 \frac{M}{\bar{\phi}}=-\frac{8\pi}{\bar{\phi}}p_{\rm T},
\end{equation}
\begin{equation}
\ddot{\bar{\phi}}+3H\dot{\bar{\phi}}+\frac{\omega'(\bar{\phi})}{2\omega(\bar{\phi})+3}\dot{\bar{\phi}}^2+\frac{2M}{2\omega(\bar{\phi})+3}=\frac{8\pi}{2\omega(\bar{\phi})+3}(\rho_{\rm T}-3p_{\rm T})
\end{equation}
where the dot denotes derivative with respect to the physical time $t$ and $H\equiv\frac{\dot{a}}{a}$ is the Hubble rate.  $\rho_{\rm T}$ is the total density of the dust and radiation in the universe and $p_{\rm T}$ is the total pressure. Here we have assumed a spatially flat FLRW universe, although the waveforms in the previous subsection apply to a general FLRW universe. 

If $\bar{\phi}$ does not evolution with the expansion of the universe, then the above equations reduce to that in GR. The last equation requires $\omega(\bar{\phi})\to \infty$ for consistency.
In this limit, we have $\tilde{g}=K=1$, $\alpha=K_1 =0=h_b$, and $D_G=D_S=D_L$. The above waveforms are reduced to those in GR.

Given the functional expression of $\omega(\phi)$, the background evolution equations can be solved to obtain the distance-redshift relations $D_G(z)$ and $D_S(z)$. When $\omega(\phi)$ is a constant, these equations have been solved in \cite{2019ApJ...886L...6S}.
When the cosmological redshift of the GW source is negligible, $D_G=D_S=r$ and the above waveforms are reduced to those in the local wave zone in section \ref{wavelocal}.

\section{Bias due to the inconsistent waveforms}\label{estbias}
In recent years, there has been an increasing amount of literature on standard sirens in modified gravity theories \cite{PhysRevD.99.083504,PhysRevD.102.044036,PhysRevD.103.064075,Baker_2021,Belgacem_2019,Dalang_2019,PhysRevD.98.023510,PhysRevD.100.044041,PhysRevLett.124.061101,Belgacem2019,PhysRevD.103.104059,Amendola2018}.
However, the waveforms used in all the previous studies, such as Eq. (14) in \cite{PhysRevD.99.083504} and Eq. (12) in \cite{PhysRevD.103.104059}, consider the modification of the cosmological distance but do not take into account the corrections in the amplitude and the phase and the extra polarization(s). That is, they have ignored the modifications due to wave generation and only consider the modification in wave propagation.
 In the following, we will analyze the bias due to the inconsistent waveforms.
In GW data analysis, it is a common practice  to use the frequency domain GW waveforms. The Fourier transforms for the waveforms of the three polarizations can be computed via the stationary phase approximation \cite{PhysRevD.98.083023,PhysRevD.102.124035}. 
\begin{align}\label{hpf}
\tilde{h}_+(f)=&-\left[1+\alpha\{\frac13(1-2s_1)(1-2s_2)-\frac{1}{12}\Gamma^2-\frac{5}{48}\mathcal{S}^2\eta^\frac25(\pi f G M_z)^{-\frac23}\}\right] \nonumber \\
&\times\left(\frac{5\pi}{24}\right)^\frac12\frac{(GM_z)^2}{D_G}\frac{1+\cos^2\iota}{2}(\pi f GM_z)^{-\frac76}e^{i\psi_+},\\
\tilde{h}_\times(f)=&-\left[1+\alpha\{\frac13(1-2s_1)(1-2s_2)-\frac{1}{12}\Gamma^2-\frac{5}{48}\mathcal{S}^2\eta^\frac25(\pi f G M_z)^{-\frac23}\}\right] \nonumber \\
&\times\left(\frac{5\pi}{24}\right)^\frac12\frac{(GM_z)^2}{D_G}\cos\iota~(\pi f GM_z)^{-\frac76}e^{i\psi_\times},
\end{align}
and
\begin{equation}
\tilde{h}_b(f) = \tilde{h}_{b1}(f)+\tilde{h}_{b2}(f),
\end{equation}
\begin{align}
\tilde{h}_{b1}(f)=& \left[1-\alpha(\frac{1}{12}\Gamma^2+\frac{5}{48}\mathcal{S}^2\eta^\frac25(2\pi f G M_z)^{-\frac23})\right] \nonumber \\
&\times\alpha \left(\frac{5\pi}{12}\right)^\frac12 \frac{(GM_z)^2}{D_S}\mathcal{S}\eta^\frac15\sin\iota(2\pi f G M_z)^{-\frac32}e^{i\psi_{b1}},\\
\tilde{h}_{b2}(f) = &-\left[1+\alpha\{\frac13(1-2s_1)(1-2s_2)-\frac{1}{12}\Gamma^2-\frac{5}{48}\mathcal{S}^2\eta^\frac25(\pi f G M_z)^{-\frac23}\}\right] \nonumber \\
& \times\alpha \left(\frac{5\pi}{96}\right)^\frac12 \frac{(GM_z)^2}{D_S}\Gamma\sin^2\iota(\pi f G M_z)^{-\frac76}e^{i\psi_{b2}} 
\end{align}

with phases
\begin{align}
\psi_+(f)=&2\pi f t_z-\psi_c-\frac{\pi}{4}+\frac{3}{128}(\pi f G M_z)^{-\frac53}\left[1-\alpha(\frac23(1-2s_1)(1-2s_2)+\frac16 \Gamma^2)\right] \nonumber \\ 
&-\frac{5}{1792}\alpha \mathcal{S}^2\eta^{\frac25}(\pi f G M_z)^{-\frac73} , \\    
\psi_\times(f) =& \psi_+ +\frac{\pi}{2}, \\
\psi_{b1}(f)=& 2\pi f t_z-\frac{\psi_c}{2}-\frac{\pi}{4}+\frac{3}{256}(2\pi f G M_z)^{-\frac53}\left[1-\alpha(\frac23(1-2s_1)(1-2s_2)+\frac16 \Gamma^2)\right] \nonumber \\
&-\frac{5}{3584}\alpha \mathcal{S}^2\eta^{\frac25}(2\pi f G M_z)^{-\frac73} , \\
\psi_{b2}(f)=&\psi_+(f). \label{psib2f}
\end{align}
When the redshift is negligible, the above tensor polarization waveforms are consistent with Eqs. (91)-(93) in \cite{PhysRevD.102.124035} and the order $O(\frac{1}{\bar{\omega}})$ terms of scalar waveforms are consistent with Eqs. (52), (53), and (55) in \cite{PhysRevD.95.124008}.

In the Brans-Dicke theory, the signal received by a GW detector is given by the following response function \cite{Poisson2014}
\begin{equation}
h_{\rm BD}(t) = F_+ h_+ + F_\times h_\times + F_b h_b,
\end{equation}
where the detector antenna pattern functions $F_A(A = +,\times, b)$ depend on the geometry of the detector and the sky location of the GW source. For the explicit expressions of the antenna pattern functions of the Einstein Telescope, see Appendix C in \cite{PhysRevD.95.124008}. 

The GW waveforms used in the previous studies on standard sirens in modified gravity theories are inconsistent in the sense that these works have ignored   modifications  from wave generation \cite{PhysRevD.99.083504,PhysRevD.102.044036,PhysRevD.103.064075,Baker_2021,Belgacem_2019,Dalang_2019,PhysRevD.98.023510,PhysRevD.100.044041,PhysRevLett.124.061101,Belgacem2019,PhysRevD.103.104059,Amendola2018}, which is equivalent to using the following frequency domain response function
\begin{equation}\label{hin}
\tilde{h}_{\rm In}(f) =F_+ \tilde{h}_+^{\rm In}(f)+ F_\times \tilde{h}_\times^{\rm In}(f),
\end{equation}
where
\begin{align}
\tilde{h}_+^{\rm In}(f)=&-\left(\frac{5\pi}{24}\right)^\frac12\frac{(GM_z)^2}{D_G}\frac{1+\cos^2\iota}{2}(\pi f GM_z)^{-\frac76}e^{i\psi_+^{\rm GR}}, \\
\tilde{h}_\times^{\rm In}(f)=&-\left(\frac{5\pi}{24}\right)^\frac12\frac{(GM_z)^2}{D_G}\cos\iota~(\pi f GM_z)^{-\frac76}e^{i\psi_\times^{\rm GR}}
\end{align}
with phases
\begin{align}
\psi_+^{\rm GR}(f)=&2\pi f t_z-\psi_c-\frac{\pi}{4}+\frac{3}{128}(\pi f G M_z)^{-\frac53}, \\
\psi_\times^{\rm GR}(f) =& \psi_+^{\rm GR}(f) +\frac{\pi}{2}.
\end{align}
It is convenient to rewrite the response function in the following form 
\begin{equation}\label{hin2}
\tilde{h}_{\rm In}(f)=Q M_z^\frac56 f^{-\frac76}e^{i\psi_+^{\rm GR}}
\end{equation}
with \footnote{Actually, $Q$ has a phase factor which has been absorbed by $\psi_c$.}
\begin{equation}\label{Q}
Q = -\frac{1}{\pi^{2/3}D_G }\left(\frac{5}{96}\right)^\frac12\sqrt{F_+^2(1+\cos^2\iota)^2+4 (F_\times \cos\iota)^2}.
\end{equation}

It is obvious that the inconsistent response function $\tilde{h}_{\rm In}(f)$ only modifies the response function in GR by replacing $D_L$with $D_G$. Note that the response function $\tilde{h}_{\rm In}(f)$ can also be obtained by setting the sensitivities of the compact objects as the ones for black holes in Eqs. \eqref{hpf}-\eqref{psib2f} to  $s_1 = s_2 = \frac12$. This is because binary black hole systems emit no scalar waves in the Brans-Dicke theory. For the binary neutron star systems, the components have similar sensitivities and the scalar dipole radiation will be suppressed. Therefore, we will focus on the GWs emitted by asymmetric black hole-neutron star (BH-NS) binaries. Since the Einstein Telescope is more sensitive to the difference between the inconsistent waveforms and the consistent waveforms, we now consider the GW observations by the Einstein Telescope and estimate the typical magnitude of the bias due to the inconsistent GW waveforms, using the method proposed by Cutler and Vallisneri \cite{PhysRevD.76.104018}.  The waveforms in the previous section apply to a general coupling function $\omega(\phi)$, while we now set  $\omega(\phi)$ to  a constant $\omega(\phi) = \omega_{\rm BD}$ for ease of numerical calculations in this section.
In this case, $D_S = D_G$ and $\alpha = \frac{1}{2\omega_{\rm BD}+3}$. 

\subsection{Method}
Consider the measured detector data
\begin{equation}
s(t) = h_{\rm BD}(t;p_{\rm tr}) + n(t), 
\end{equation}
where $n(t)$ is the detector noise and $p_{\rm tr}$ denotes the true GW source parameters collectively. If we use the inconsistent response function $h_{\rm In}(t)$ to estimate the parameters, we determine the best-fit value $p_{\rm bf}$ by minimizing the noise-weighted inner product \cite{PhysRevD.76.104018}
\begin{equation}
 (s - h_{\rm In}|s - h_{\rm In}).
\end{equation} 
For any two given signals $h_1(t)$ and $h_2(t)$, the inner product is given by
\begin{equation}
(h_1|h_2) = 4 \Re \int_0^\infty \frac{\tilde{h}_1(f)\tilde{h}_2^*(f)}{S_n(f)} df
\end{equation}
where $\tilde{h}_1(f)$ and $\tilde{h}_2(f)$ are the Fourier transforms of $h_1(t)$ and $h_2(t)$; $S_n(f)$ is the one-side noise power spectral density (PSD). The noise PSD of the Einstein Telescope is assumed to be the ET-D model \cite{Hild_2011}.
Then, the best-fit parameters $p_{\rm bf}$ satisfy \cite{PhysRevD.76.104018}
\begin{equation}
(\partial_j h_{\rm In}(p_{\rm bf})|s - h_{\rm In}(p_{\rm bf})) = 0,
\end{equation}
where $\partial_j h_{\rm In}= \partial h_{\rm In}/ \partial p^j$ and $p^j$ denotes the GW source parameters. It can be seen from Eq. \eqref{hin2} that the response function $h_{\rm In}$ depends on four parameters $p^j = (\ln Q, \ln M_z, f_0 t_z, \psi_c)$.
From the above equation, to the first order in the error 
\begin{equation}
\Delta p^i \equiv p_{\rm bf}^i - p_{\rm tr}^i,
\end{equation}
we have \cite{PhysRevD.76.104018}
\begin{equation}\label{dp}
\Delta p^i =\Delta_n p^i + \Delta_{\rm th}p^i,
\end{equation}
where
\begin{align}
\Delta_n p^i =& (\Lambda^{-1})^{ij}(\partial_j h_{\rm In}(p_{\rm bf})|n),\\
\Delta_{\rm th}p^i =&(\Lambda^{-1})^{ij}(\partial_j h_{\rm In}(p_{\rm bf})|h_{\rm BD}(p_{\rm tr})-h_{\rm In}(p_{\rm tr})) \label{dthp}
\end{align}
with $(\Lambda^{-1})^{ij}$  the inverse of the Fisher matrix
\begin{equation}\label{Fisher}
\Lambda_{ij}(p_{\rm bf}) \equiv (\partial_i h_{\rm In}(p_{\rm bf})|\partial_j h_{\rm In}(p_{\rm bf})).
\end{equation}
It can be seen that $\Delta_n p^i$ is the statistical error  due to the noise and $\Delta_{\rm th}p^i$ is the theoretical bias  due to the waveform inconsistency. The standard deviation of the statistical errors is given by \cite{PhysRevD.46.5236,PhysRevD.49.2658}
\begin{equation}\label{std}
\delta_n p^i = \sqrt{\langle \Delta_n p^i \Delta_n p^i \rangle} =\sqrt{(\Lambda^{-1})^{ii}}.
\end{equation}
Here `$\langle  \rangle$' represents ensemble average  and no summation is implied by the repeated index. 
We will now  compare the standard deviation $\delta_n p^i$ and the theoretical bias $\Delta_{\rm th}p^i$ in GW observations using the PSD of the instrumental noise of  the Einstein Telescope.

Substituting  the response function \eqref{hin2} into Eq. \eqref{Fisher} and using Eq. \eqref{std}, we obtain the standard deviations of the amplitude parameter and the redshifted chirp mass for GW observations by the Einstein Telescope
\begin{align}
\delta_n \ln Q =& \frac{1}{\rho},\label{dq}\\
\delta_n \ln M_z =& 1.3\times10^{-5}\left(\frac{M_z}{M_\odot}\right)^{\frac53}\frac{1}{\rho}, \label{dm}
\end{align}
where $\rho$ is the signal-to-noise ratio (SNR) given by
\begin{equation}
\rho = \sqrt{ ( h_{\rm In}|h_{\rm In})}.
\end{equation}

To obtain the theoretical bias, we need the difference between the two response functions
\begin{equation}\label{wavediff}
\tilde{h}_{\rm BD}(f)-\tilde{h}_{\rm In}(f) = (\Delta A + i \Delta\psi)\tilde{h}_{\rm In}(f)+ F_b \tilde{h}_{b2},
\end{equation}
where the amplitude correction to the tensor polarizations is
\begin{equation}
\Delta A = \alpha(A_1+A_2(\pi f G M_z)^{-\frac23})
\end{equation}
with 
\begin{equation}
A_1 = \frac13 (1-2s_1)(1-2s_2)-\frac{1}{12}\Gamma^2,\qquad A_2 = -\frac{5}{48}\mathcal{S}^2\eta^\frac25
\end{equation}
and the phase correction to the tensor polarizations is
\begin{equation}
\Delta \psi = \alpha (\psi_1 (\pi f G M_z)^{-\frac53}+\psi_2 (\pi f G M_z)^{-\frac73})
\end{equation}
with
\begin{equation}
\psi_1 = -\frac{3}{128}(\frac23 (1-2s_1)(1-2s_2)+\frac{1}{6}\Gamma^2),\qquad \psi_2 = -\frac{5}{1792}\mathcal{S}^2\eta^\frac25.
\end{equation}
Here $F_b \tilde{h}_{b2}$ denotes the contribution from the breathing polarization. Since $\tilde{h}_{b1}$ will contribute an oscillatory term to the integrand of the inner product in Eq. \eqref{dthp}, its contribution can be discarded. The breathing polarization contribution can be rewritten as 
\begin{equation}
F_b \tilde{h}_{b2} = \alpha B_2 \Gamma \tilde{h}_{\rm In}(f)
\end{equation}
with 
\begin{equation}
B_2 = \frac{F_b\sin^2\iota}{F_+(1+\cos^2\iota)+2iF_\times\cos\iota}.
\end{equation}
Note that we  keep only terms to order $O(\alpha)$ in the difference.
Using Eq. \eqref{dthp} and the difference between the response functions, we obtain the theoretical biases for GW observations by Einstein Telescope
\begin{align}
\Delta_{\rm th}\!\ln Q =& \alpha[A_1 + 247 A_2 \left(\frac{M_z}{M_\odot}\right)^{-\frac23}+21.3 \psi_1 +1.26\times 10^4~\psi_2\left(\frac{M_z}{M_\odot}\right)^{-\frac23} +\Re[B_2]\Gamma],\label{Dq}\\
\Delta_{\rm th}\!\ln M_z =& -\alpha[25.6\psi_1+1.51\times10^4~\psi_2\left(\frac{M_z}{M_\odot}\right)^{-\frac23}]. \label{Dm}
\end{align}
From here we see that the theoretical biases are proportional to $\alpha = \frac{1}{2\omega_{\rm BD}+3}$ and independent of SNR. The amplitude bias $\Delta_{\rm th}\!\ln Q$ has contributions from the extra polarization as well as the amplitude and phase corrections of the tensor polarizations. But the mass bias $\Delta_{\rm th}\!\ln M_z$  has only contributions from the phase correction of the tensor polarizations.

\subsection{Result}

We now apply Eqs. \eqref{dq} \eqref{dm} \eqref{Dq} and \eqref{Dm} to the BH-NS binaries which can be detected by Einstein Telescope. We choose masses and redshifts of the GW sources to be that of the BH-NS  binaries detected by LIGO/Virgo.
Up to now, among all the compact binaries detected by LIGO/Virgo, there are several  BH-NS  candidates. Two of the candidate BH-NS binaries (GW191219 and GW200115) have a secondary with mass below $2M_\odot$, which can be confidently interpreted as a NS \cite{2021arXiv211103606T}. Their source parameters are listed in Table \ref{para}. Because the Einstein Telescope's noise PSD will be lower and its sensitive frequency band wider than that of the advanced LIGO, we set the SNR  in Eqs. \eqref{dq} and \eqref{dm} to be ten times of the SNR detected by LIGO/Virgo \footnote{We have generated one hundred GW191219-like sources, with the source masses and source distances set to be those of GW191219, but the sky positions and orbital orientations randomly set, and calculated the SNR detected by the Einstein Telescope using the package \textsc{gwbench} \cite{Borhanian_2021}. For more than 30 GW191219-like sources, the SNR detected by the Einstein Telescope is greater than ten times the SNR of GW191219 detected by LIGO/Virgo. We have tested the GW200115-like sources in the same way and obtained the similar result.  }. The sensitivity of NS is set to  0.2 \cite{1992ApJ...393..685Z}. 

Note that for a GW binary source with cosmological redshift $z = 0.06$, the redshift is high enough that the local wave zone waveforms are not applicable, and it is necessary to use waveforms \eqref{otz}-\eqref{hb2z} to study this GW source.
The formulae of the statistical errors and the theoretical biases in the previous subsection are applicable  to any source with a SNR relatively
higher (SNR>20). The assumption of the Fisher matrix analysis (Gaussian approximation) will become invalid  for low SNR events.
We leave a more comprehensive analysis based on the population properties of the GW sources as a future work
\cite{2021arXiv211103634T}.

\begin{table}[h]
\caption{ Selected source parameters of BH-NS candidates detected by LIGO/Virgo. The columns show source component masses $m_A$, redshift $z$, and the network matched-filter SNR.  These data are taken from Table IV of \cite{2021arXiv211103606T}.}\label{para}
\begin{tabular}{ccccc}
\hline\hline
 & $m_1/M_\odot$ & $m_2/M_\odot$ & $z$ & SNR \\ 
\hline
GW191219 & 31.1 & 1.17 & 0.11 & 9.1\\
\hline
GW200115 & 5.9 & 1.44 & 0.06 & 11.3\\
\hline\hline
\end{tabular}
\end{table}

The numerical results of the theoretical biases and the statistical errors are shown in Table \ref{result}. The amplitude bias
$\Delta_{\rm th}\!\ln Q$ depends on $\Re[B_2]$ which is related to the viewing angle $\iota$ and the antenna pattern functions. We generate $10^6$ samples of $\Re[B_2]$ by randomly choosing the sky location and the orientation of the angular momentum of binary system. The median of the absolute value of these samples is 0.11. From Table \ref{result} we see that, in general, the dominant contribution to  amplitude bias $\Delta_{\rm th}\!\ln Q$ is from $\psi_2$ which originates from the scalar dipole radiation, and the dominant contribution to  mass bias $\Delta_{\rm th}\!\ln M_z $ is also from the phase correction of scalar dipole radiation.

\begin{table}[h]
\caption{Theoretical biases and statistical errors in GW observation by the Einstein Telescope. Here $Q_i$ represents the contribution of the $i$-th term on the right-hand side of Eq. \eqref{Dq} and $M_i$ represents the contribution of the $i$-th term on the right-hand side of Eq. \eqref{Dm}.  }\label{result}
\begin{tabular}{c*{11}{c}}
\hline\hline
 & $(\Delta_{\rm th}\!\ln Q) /\alpha$ & $Q_1$ & $Q_2$ &$Q_3$ & $Q_4$ & $Q_5$ & $(\Delta_{\rm th}\!\ln M_z) /\alpha$ & $M_1$ & $M_2$ & $\delta_n\! \ln Q$ & $\delta_n\! \ln M_z$\\ 
\hline
GW191219-like &  $-0.56+ 0.58 \Re[B_2]$ & $-0.028$ & $-0.21$ & $-0.028$ & $-0.29$ & $ 0.58 \Re[B_2]$ & 0.38 & 0.033& 0.35 & 0.011 & $1.9\times 10^{-6}$\\
\hline
GW200115-like &  $-1.4+ 0.48 \Re[B_2]$ & $-0.019$ & $-0.59$ & $-0.019$ & $-0.81$ & $ 0.48 \Re[B_2]$ & 0.99 & 0.023 & 0.97 & $8.8\times 10^{-3}$ & $5.5\times 10^{-6}$\\
\hline\hline
\end{tabular}
\end{table}

\begin{figure}
\begin{center}
\includegraphics[width=8cm]{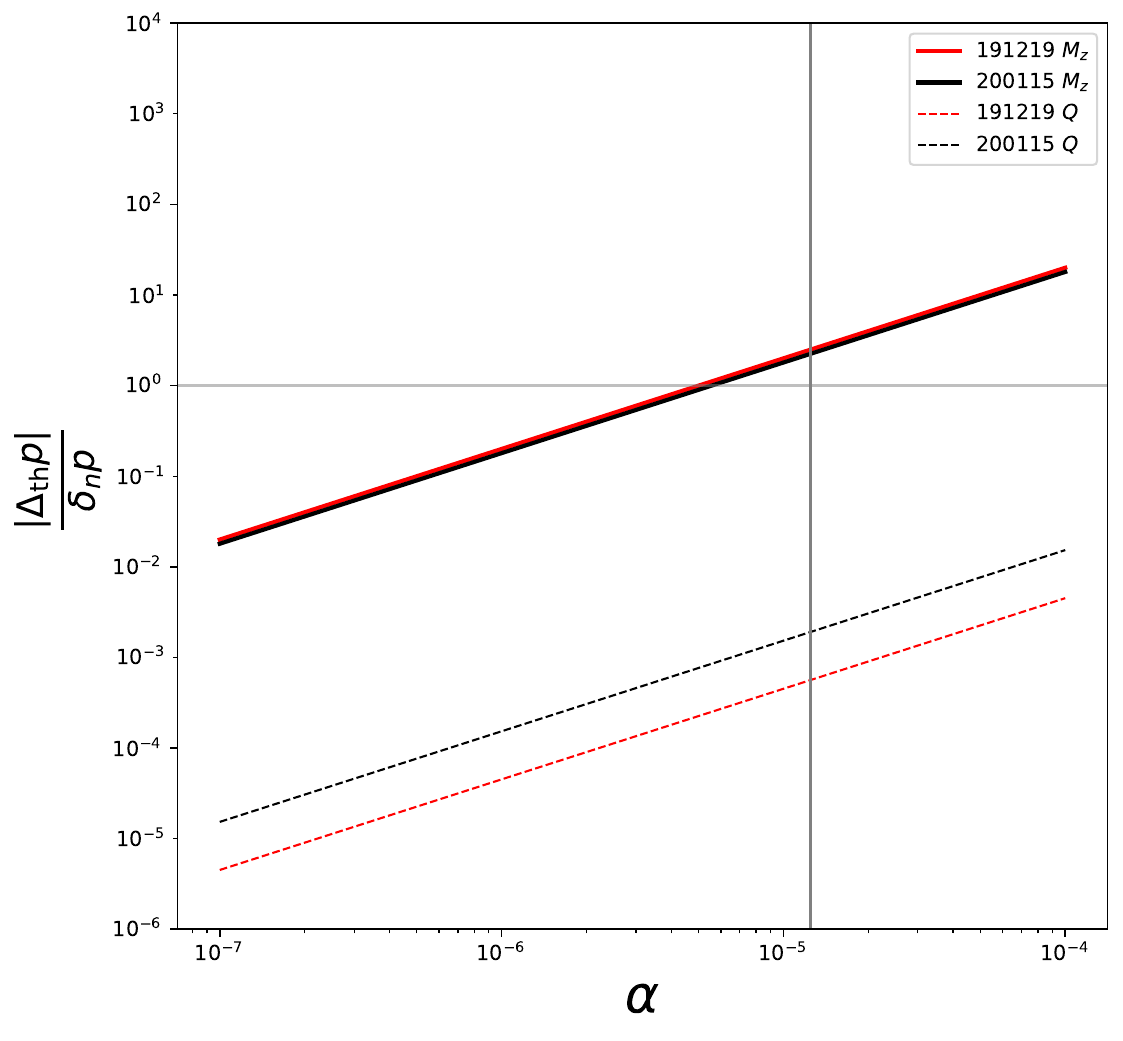}
\caption{The ratios of the theoretical bias to the statistical error increases with the coupling constant $\alpha$. Colors represent different BH-NS binaries. The solid lines and the dashed lines represent $\frac{|\Delta_{\rm th}\ln M_z|}{\delta_n \ln M_z}$ and $\frac{|\Delta_{\rm th}\ln Q|}{\delta_n \ln Q}$, respectively. The vertical  line indicates the upper bound on $\alpha$ imposed by the Cassini spacecraft.
}\label{biasfig}
\end{center}
\end{figure}

Then we compare the theoretical bias and the statistical error. The ratio of the theoretical bias to the statistical error is shown in Fig. \ref{biasfig}. To obtain the result, $\Re[B_2]$ is set to the  median 0.11. The ratio increases with the coupling constant $\alpha$. The positive coupling constant $\alpha$ has be constrained to be less than $1.3\times 10^{-5}$ by the Cassini spacecraft \cite{PhysRevD.85.064041}. In this paper, we set the interesting range of $\alpha$ to $10^{-7}-10^{-4}$ (cf. FIG. \ref{biasfig}) to leave some margins and ensure that the linear approximation of the difference of the waveforms, Eq.\eqref{wavediff}, is applicable. It can be seen that, for these two BH-NS binaries, when $\alpha$ approaches the upper bound imposed by the Cassini spacecraft $\frac{|\Delta_{\rm th}\ln M_z|}{\delta_n \ln M_z}>1$ and $\frac{|\Delta_{\rm th}\ln Q|}{\delta_n \ln Q}<0.01$. Therefore, the waveform inconsistency can  affect the measurement accuracy of the redshifted chirp mass significantly in this situation. From Eq. \eqref{Q} we see that $\ln Q = -\ln D_G+\ln(\text{angles})$. If we interpret the response function \eqref{hin2} as averaging over the angels, then $\Delta\ln Q = -\Delta\ln D_G$. Because of this, the waveform inconsistency has a negligible impact on the measurement accuracy of the source distance $D_G$. This is only a preliminary study on the measurement accuracy of the distance. A more detailed and comprehensive analysis of the measurement of the distance and the Hubble constant is subject to our future work.

 % Assuming $|\Delta_{\rm th}p^i|>\delta_n p^i$, we obtain the upper bounds on $\omega_{\rm BD}$ shown in Table \ref{compare}. To obtain these results, $\Re[B_2]$ is set to the  median 0.11. However $\omega_{\rm BD}$ has been constrained to be larger than 40~000 by the Cassini spacecraft \cite{PhysRevD.85.064041}. Therefore, if $\omega_{\rm BD}$ saturates the Cassini constraint, the theoretical bias of the amplitude $Q$ is negligibly small compared with its statistical error, while the waveform inconsistency can still influence the measurement of the redshifted chirp mass significantly. From Eq. \eqref{Q} we see that $\ln Q = -\ln D_G+\ln(\text{angles})$. If we interpret the response function \eqref{hin2} as averaging over the angels, then $\Delta\ln Q = -\Delta\ln D_G$. Because of this, the waveform inconsistency has a negligible influence on the measurement of the source distance $D_G$. This is only a preliminary study on the measurement of the distance. A more detailed and comprehensive analysis of the measurement of the distance and Hubble constant is left as a future work.

% \begin{table}[h]
% \caption{ Upper bound on $\omega_{\rm BD}$ by assuming that theoretical bias is larger than the statistical error, $|\Delta_{\rm th}p^i|>\delta_n p^i$.}\label{compare}
% \begin{tabular}{ccc}
% \hline\hline
%  & $\ln Q$ & $\ln M_z$  \\ 
% \hline
% GW191219-like & 23 & $6.1\times 10^4$  \\
% \hline
% GW200105-like & 77 & $3.0\times 10^5$ \\
% \hline
% GW200115-like & 83 & $5.7\times 10^5$ \\
% \hline\hline
% \end{tabular}
% \end{table}

\section{Conclusions and Discussions}\label{condis}
We have obtained the  self-consistent   gravitational waveforms in an expanding Universe in the Brans-Dicke theory. There are three GW polarizations, i.e., the plus polarization $h_+$, the cross polarization $h_\times$, and the breathing polarization $h_b$. The waveforms of these three polarizations can be applied to  high redshift GW sources. In the GR limit ($\bar{\omega}\to  \infty$) or when the redshift of GW source is negligible ($z\ll1$), the waveforms converge to the  previously known results.

We have considered the modifications during wave generation and  propagation. Previous researches on standard sirens in modified gravity theories have adopted  inconsistent waveforms which has ignored the modifications  from wave generation; but the waveforms that focus on the wave generation effects, such as the post-Newtonian waveforms,  apply only to the local wave zone and cannot be used directly to study GW sources with non-negligible redshifts. Compared with GR, in wave generation, there are extra breathing polarization as well as  phase  and amplitude corrections to the tensor polarizations; in wave propagation, the electromagnetic luminosity distance $D_L$ is replaced by the gravitational distance $D_G$  and the scalar distance $D_S$. The expression of $D_G$ has been derived before, e.g., \cite{PhysRevD.102.044036}, while $D_S$ is a new result. Inconsistent waveforms can bias the source parameters measured by the GW detectors. The main contribution to parametric estimation  biases is from the phase correction of the frequency domain waveforms originating from scalar dipole radiation. This shows that the expanding of Universe will not suppress the modification of the waveforms due to the scalar dipole radiation.   We found that, when the coupling constant $\alpha$ approached  the Cassini upper bound, in the detection of BH-NS binaries (GW191219 and GW200115) by the Einstein Telescope,  the inconsistent waveforms would bias the measurement of redshifted chirp mass significantly but had negligible impact on the measurement of the distance.

The solution to the geometric optics equations, Eq. \eqref{Ap}, is a general result which can be applied to the GWs emitted by any isolated sources, such as a binary with a precessing orbital plane \cite{PhysRevD.95.104038},  an extreme
mass ratio inspiral  system \cite{PhysRevD.85.102003}, a  triple system \cite{PhysRevD.100.083012}, etc. The method using local wave zone to match wave generation and wave propagation can also be used in other theories, such as massive Brans-Dicke theory \cite{PhysRevD.102.124035}, reduced Horndeski theory \cite{PhysRevD.103.064075}, Chern-Simons theory \cite{PhysRevD.85.064022,*PhysRevD.93.029902}, etc. 
It is interesting to consider  using this match procedure to combine the parametrized post-Einsteinian (ppE) framework \cite{PhysRevD.98.084042,*PhysRevD.101.109902} with the generalized GW propagation (gGP) framework \cite{PhysRevD.97.104037,*PhysRevD.97.104038,*PhysRevD.99.104038}. The ppE framework mainly focuses on wave generation in modified gravity theories while the gGP framework deals with the wave propagation. 

In addition to the modification of GW waveforms, modified gravity can influence the cosmological background evolution \cite{Belgacem2019,PhysRevD.100.044041}. It is necessary to include all these effects to study standard sirens in the Brans-Dicke theory. We leave this for future work.

\begin{acknowledgments}
T. L. is supported by the National Natural Science Foundation of China (NSFC) Grant No. 12003008 and the China Postdoctoral Science Foundation Grant No. 2020M682393. 
Y.W. gratefully acknowledges support from the National Key Research and Development Program of China (No. 2022YFC2205201 and No. 2020YFC2201400), the NSFC under Grants No. 11973024, Major Science and Technology Program of Xinjiang Uygur Autonomous Region (No. 2022A03013-4), and Guangdong Major Project of Basic and Applied Basic Research (Grant No. 2019B030302001).
W.Z. is supported by NSFC Grants No. 11773028, No.
11633001, No. 11653002, No. 11421303, No. 11903030,
the Fundamental Research Funds for the Central Universities, and the Strategic Priority Research Program of the
Chinese Academy of Sciences Grant No. XDB23010200. The authors thank the anonymous referees for
helpful comments and suggestions.
\end{acknowledgments}

% \bibliography{sirenBD,mBDwaveform}


\end{document}